\documentclass[a4paper,11pt]{article}
\usepackage{jheppub}
\usepackage{graphicx}
\usepackage{xcolor}
\begin{document}
\title{\boldmath Cosmic strings in conformal gravity}

\author[a,1]{R. J. Slagter\note{Corresponding author.}}
\affiliation[a]{ASFYON, Astronomisch Fysisch Onderzoek Nederland and  Department of Physics, University of Amsterdam, The Netherlands }
\author[b]{and C. L. Duston}
\affiliation[b]{Physics Department, Merrimack College, N Andover, MA, USA}

\emailAdd{info@asfyon.com}
\emailAdd{dustonc@merrimack.edu}
\abstract{
We investigate the spacetime of a spinning cosmic string in conformal invariant gravity, where the interior consists of a gauged scalar field.
We find  exact solutions of the exterior of a stationary spinning cosmic string, where we  write the metric as $ g_{\mu\nu}=\omega^2\tilde g_{\mu\nu}$, with $\omega$ a dilaton field which contains all the scale dependences.
The "unphysical" metric $\tilde g_{\mu\nu}$ is related to the $(2+1)$-dimensional Kerr spacetime.
The equation for the angular momentum $J$ decouples, for the vacuum situation as well as for global strings, from the other field equations and delivers a kind of spin-mass relation.
For the most realistic solution, $J$ falls off as $\sim\frac{1}{r}$ and  $\partial_r J \rightarrow 0$ close to the core. The spacetime is Ricci flat. The formation of closed timelike curves can be pushed to space infinity for suitable values of the parameters and the violation of the weak energy condition can be avoided.
For the interior, a numerical solution is found. This solution can easily be matched  at the boundary on the exterior exact solution by special choice of the parameters of the string.
It turns out, as expected from the "holographic"  principle, that the exact solution of the exterior is equivalent with the warped five-dimensional brane world model, with only a cosmological constant in the bulk.
This example shows the power of conformal invariance to bridge the gap between general relativity and quantum field theory.
}

\maketitle
\flushbottom
\section{Introduction}\label{Intro}	
In general relativity theory (GRT) one can construct solutions which are related to real physical objects. The most famous one is the black hole solution. One now believes
that in the center of many galaxies there is a rotating super-massive black hole, the Kerr black hole. Because there is an axis of rotation, the Kerr solution is a member of the family of the axially symmetric solutions of the Einstein equations\cite{Steph:2009}.
A legitimate question is if there are other axially or cylindrically symmetric asymptotically flat, but non-static solutions of the equations of Einstein with a classical or non-classical matter distribution and with correct asymptotical behavior, just as the Kerr solution. Many attempts are made, such as the Weyl-, Papapetrou- and Van Stockum solution\cite{Islam:1985}. None of these attempts result in physically acceptable solution.
Often, these solutions possess closed timelike curves (CTC's).  The possibility of the formation of CTC's in GRT seems to be an obstinate problem to solve in GRT. At first glance, it seems possible to construct in GRT causality violating solutions. CTC's suggest the possibility of time-travel with its well-known paradoxes.
The first spacetime in which CTC's would exist was that of G\"odel\cite{Godel:1949}. It represents a homogeneous rotating universe without expansion, so in conflict with standard cosmological models. Nevertheless, even Einstein was surprised by  the G\"odel solution. For a mathematical treatment of the  G\"odel universe, we refer to the book of Hawking and Ellis\cite{Hawk:1973} and Barrow\cite{Barrow:2004}.
Although most physicists believe that Hawking's chronology protection conjecture holds in our world, it can be alluring  to investigate the mathematical underlying arguments of the formation of CTC's. There are several spacetimes that can produce CTC's. Most of them can easily characterized as un-physical. The problems are, however, more deep-seated in the vicinity of a (spinning) cosmic string or in the so-called Gott-spacetime. These
cosmic strings models gained much attention the last decades\cite{stockum:1937,hiscock:1985,gott:1985,deser2:1992,jensen:1992,slagter1:1996,slagter1a:2002,bonner:2002}.
Two cosmic strings, approaching each other with high velocity, could produce CTC's. If an advanced civilization could manage to make a closed loop around this Gott pair, they will be returned to their own past.
However, the CTC's will never arise spontaneously from regular initial conditions through the motion of spinless cosmons ( “Gott’s pair”):  there  are boundary conditions that has CTC's also at infinity  or at an initial  configuration\cite{thooft2:1992,thooft3:1993}. If it would be possible to fulfil  the CTC condition at $t_0$ then at sufficiently large times  the cosmons will have evolved so far apart that the CTC's would disappear.
The chronology protection conjecture seems to be saved for the Gott spacetime.
There  are still some unsatisfied aspects around spinning cosmic strings.
If the cosmic string has a finite dimension, one needs to consider the coupled field equations, i.e., besides the Einstein equations, also the scalar and gauge field equations\cite{lag1:1985,lag2:1989,garf1:1985}.
It came as a big surprise that there exists a vortex-like solution in GRT comparable with the magnetic flux lines in type II superconductivity\cite{Niels:1973}. Many of the features of the Nielsen-Olesen vortex solution and superconductivity will survive in the self-gravitating situation. These vortex lines occur as topological defects in an abelian U(1) gauge model, where the gauge field is coupled to a charged scalar field.
It can  easily be established that the solution must be cylindrically symmetric, so independent of the z-coordinate and the energy per unit length along the z-axis is finite.
There are two types, local (gauged) and global cosmic strings. We are mainly interested in local cosmic strings, because in a gauge model,  strings were formed during a local symmetry breaking and so have a sharp cutoff in energy, implying no long range interactions.
Spinning cosmic string solutions can cause serious problems when CTC's are formed which are not hidden behind a horizon, as is the case for the Kerr metric.
One can "hide" the presence of the spinning string by suitable coordinate transformation in order to get the right asymptotic behavior. One obtains then a helical structure of time, not desirable.
Further, it is not easy to match the interior on the vacuum exterior and to avoid the violation of the weak energy condition (WEC)\cite{janca:2007,Krisch:2003}.

In this manuscript we will consider the spinning string in conformal invariant gravity.
Conformal invariance (CI) was originally introduced by Weyl\cite{Weyl:1918}. His original idea was to introduce a new kind of geometry, in relation to a unified theory of gravitation and electromagnetism. This approach was later abandoned with the birth of modern gauge field theories.
Quite recently the Anti-deSitter/Conformal field-theory (AdS/CFT)   correspondence renewed the interest in conformal gravity. AdS/CFT is a conjectured relationship between two kinds of physical theories. AdS spaces  are used in theories of quantum gravity while  CFT  includes theories similar to the Yang–Mills theories that describe elementary particles\cite{Mald:1998}.
The main conclusion in this model was that expectation values of the metric in five dimensional AdS spacetime are equal, at tree level, to expectation values computed using  four dimensional conformal gravity. It conjectures an equivalence between string theory in AdS space and a CFT on its boundary.
AdS/CFT correspondence is a prime example of the  "holographic" principle, first proposed by 't Hooft\cite{thooft1:1993}.

It is now believed that CI can help us move a little further along the road to quantum gravity.
Conformal invariance in GRT considered as exact at the level of the Lagrangian but spontaneously broken just as in the case of the Brout-Englert-Higgs mechanism (BEH) in standard model of particle physics, is an approved alternative for disclosing the small-distance structure when one tries to describe quantum-gravity problems\cite{thooft0:2010,thooft4:2011}  and can also be used to model scale-invariance in the CMB\cite{bars:2014}.
It still is a controversial alternative method to describe canonical quantum gravity, because one is saddled  with serious anomalies\cite{thooft4:2011,thooft0:2010,thooft5:2010,thooft6:2015}. The key problem is perhaps the handling of asymptotic flatness of isolated systems in GRT, specially when they radiate and the generation of the metric $g_{\mu\nu}$ from at least Ricci-flat spacetimes. Close to the Planck scale one should like to have Minkowski spacetime and somehow  curvature must  emerge. Curved spacetime will inevitable enter the field equations on small scales.
The first task is then to construct a Lagrangian, where spacetime and the fields defined on it, are topological regular and physical acceptable  in the non-vacuum case. This can be done by considering the scale factor( or warp factor in higher-dimensional models) as a dilaton field besides the conformally coupled  scalar field.
It is known since the 70s\cite{parker:2009}, that quantum field theory combined with Einstein's gravity  runs into serious problems. The Einstein-Hilbert(EH) action is non-renormalizable and it gives rise to intractable divergences at loop levels.
On very small scales, due to quantum corrections to GRT, one must modify Einstein's gravity by adding higher order terms in the Lagrangian like  $R^2, R_{\mu\nu}R^{\mu\nu}$ or $R_{\mu\nu\sigma\tau}R^{\mu\nu\sigma\tau}$ (or combinations of them).
However, serious difficulties arise in these higher-derivative models, for example,  the occurrence of massive ghosts which cause unitary problems.
A next step is then to disentangle the functional integral over the dilaton field from the ones over the metric fields and matter fields.
Moreover, it is desirable that all beta-functions of the matter lagrangian, in combination with the dilaton field, disappear in order to fix  all the coupling constants of the model.
Further, conformal invariance of the action with matter fields implies that the trace of the energy-momentum tensor is zero. A theory based on a classical "bare" action which is conformally invariant, will lose it in quantum theory as a result of renormalization and the energy-momentum tensor acquires a non-vanishing trace ( trace anomaly). In a warped 5D model, a contribution to the trace from the bulk could possible solve this problem\cite{slagter:2017}.
Recently, it was found that the  black hole complementarity between infalling and stationary observers and the information paradox, could be well described by CI\cite{thooft6:2015}.
Another interesting application  can be found in the work of Mannheim on conformal cosmology\cite{mannheim:2005,mannheim:2006}. This model could serve as an alternative approach to explain  the rotational curves of galaxies, without recourse to dark matter and dark energy (or cosmological constant).
In a former study, we investigated the gauged scalar field in context with warped spactimes\cite{slagter2:2016} and conformal invariance\cite{slagter3:2019,slagter4:2018}. New features will be encountered in the spinning case, which we will consider here.
In section 2 we introduce briefly the concept of CI in GRT. In section 3 we present  our model of a spinning cosmic string in conformal Higgs gravity. In section 4 we present the exact solution of the exterior and in section 5 we consider the connection with the warped 5D model.
\section{Introduction to conformal dilaton gravity}\label{sec2}
The use of conformal invariance in GRT is one of the possibilities to bridge the gap between  GRT and  QFT.  We know that the classical SM Lagrangian is lacking any intrinsic mass of length scales without the  Higgs mechanism. So it is quite natural to consider also GRT without any mass or length scale  before a kind of symmetry breaking, i.e., the conformal symmetry breaking.
A theory is  conformal invariant at the classical level, if its action is invariant under the conformal group. This is a local symmetry when the metric is dynamical ( in GRT) and global in Minkowski ( in QFT).
It will be clear that conformal invariance is broken (anomalously) when one approaches the quantum scale and renormalization procedures come into play.
A conformal transformation in GRT can be written as a mapping of a manifold ${\cal M}$
\begin{equation}
g_{\mu\nu}({\bf x})\rightarrow \Omega({\bf x})^2g_{\mu\nu}({\bf x}),\label{eq2-1}
\end{equation}
which preserves angles on the manifold. This is not, in general, a diffeomorphism or isometry on ${\cal M}$\cite{wald:1984}.
If $\Omega$ is strictly a constant, this transformation represents only a change in scales.
For Lorentzian spacetimes $({\cal M},g)$ and $({\cal M},\Omega^2 g)$ have the same causal structure.

Let us write
\begin{equation}
g_{\mu\nu}({\bf x})=\omega({\bf x})^2\tilde g_{\mu\nu}({\bf x})\label{eq2-1}.
\end{equation}
$\tilde g_{\mu\nu}$ is sometimes called the "un-physical" metric. If it is non-trivial, it would describe the physical phenomena without directly referring to the metascalar field  ("dilaton") $\omega$.
We will treat $\omega$ in section 3 on an equal  footing as the scalar (Higgs-) field.
If one considers $\omega$ as an independent dynamical degree of freedom\cite{thooft4:2011}, then the renormalization procedure leads to quantum counter terms in the total action proportional to the $\sim R_{\mu\nu}R_{\mu\nu}-\frac{1}{3}R^2$. When performing the functional integration, one first integrate over $\omega$ together with the matter fields and then over $\tilde g_{\mu\nu}$. One will need additional constraints on $\tilde g_{\mu\nu}$ in order to deal with the coordinate re-parametrization ambiguity.
If $\tilde g_{\mu\nu}=\eta_{\mu\nu}$, so that $\tilde R=0$, then $\omega$ will be uniquely determined. It is needed to describe clocks and rulers in the macroscopic world.
For general $\tilde g_{\mu\nu}$ we have the additional gauge freedom
\begin{eqnarray}
\tilde g_{\mu\nu}({\bf x})\rightarrow \Omega({\bf x})^2\tilde g_{\mu\nu}({\bf x}), \qquad \omega({\bf x})\rightarrow \frac{1}{\Omega({\bf x})}\omega({\bf x})\label{eq2-2}.
\end{eqnarray}
This will us then allowing  to define clocks, rulers and matter. In fact, this Weyl transformation of the metric is an exact symmetry of the action.
The conformal symmetry is described (in 4D)  by 15 parameters, representing the conformal group.
If we allow the parameters to become spacetime dependent, then we are left with only 11 local symmetries (which is correct: 10 for the metric and 1 for the dilaton).
One can construct the generators of the conformal group
\begin{eqnarray}
P_\mu =-i\partial_\mu,\quad L_{\mu\nu}=i(x_\mu\partial_\nu -x_\nu\partial_\mu ),\quad D=-i x^\mu\partial_\mu, \quad K_\mu =-i(2 x_\mu x^\nu\partial_\nu -x^\tau x_\tau \partial_\mu )\label{eq2-3}
\end{eqnarray}
The conformal group enlarge the flat space Poincar\'{e} group and is isomorphic to $SO(4,2)$.
One can apply the flat spacetime Poincar\'{e} symmetry to the curved de Sitter $SO(4,1)$ or anti-de Sitter $SO(3,2)$ spacetimes by the holographic principle.
A theory is called conformal invariant at the classical level, if its action is invariant under the conformal group.
In order to describe curvature in this context, one uses an orthonormal tetrad basis is stead of a coordinate basis and Killing vectors to describe symmetries, i.e., local Lorentz invariance. We then have 16 degrees of freedom in stead of 11. See, for example the textbook of Wald\cite{wald:1984} for details.

For compact objects, as we will see in section 3, it is convenient to impose that $\omega^2\tilde g_{\mu\nu}=\eta_{\mu\nu}$ at infinity.
\section{Conformal Higgs gravity on axially symmetric spacetimes: an example}\label{sec3}
\subsection{The conformal invariant action}\label{sec3A}
Let us now  consider, as an example, the stationary axially symmetric spacetime\cite{Steph:2009,Islam:1985}
\begin{equation}
ds^2=-e^{-2f(r)}(dt-J(r)d\varphi)^2+e^{2f(r)}\Bigl(l(r)^2d\varphi^2+e^{2\gamma(r)}(dr^2+dz^2)\Bigr)\label{eq3-1},
\end{equation}
rewritten as
\begin{equation}
ds^2=\omega(r)^2\Bigl[-(dt-J(r)d\varphi)^2+b(r)^2d\varphi^2+e^{2\mu(r)}(dr^2+dz^2)\Bigr]\label{eq3-2}.
\end{equation}
We define is this way an "un-physical" metric $\tilde g_{\mu\nu}$ by\footnote{For a more deep seated treatment of this splitting in connection with renormalization issues and effective action, we refer to\cite{thooft4:2011}.}
\begin{equation}
g_{\mu\nu}=\omega^2\tilde g_{\mu\nu}\label{eq3-3}.
\end{equation}
The model we will investigate  is given by the conformal invariant action, where we included the abelian U(1) scalar-gauge field $(\Phi, A_\mu)$
\begin{eqnarray}
{\cal S}=\int d^4x\sqrt{- \tilde g}\Bigl\{-\frac{1}{12}\Bigl(\tilde\Phi\tilde\Phi^*+\bar\omega^2\Bigr) \tilde R-\frac{1}{2}\Bigl( D_\alpha\tilde\Phi( D^\alpha\tilde\Phi)^*+\partial_\alpha\bar\omega\partial^\alpha\bar\omega\Bigr)\cr
-\frac{1}{4}F_{\alpha\beta}F^{\alpha\beta}-V(\tilde\Phi ,\bar\omega)-\frac{1}{36}\kappa^2\Lambda\bar\omega^4\Bigr\}\label{eq3-4}.
\end{eqnarray}
We parameterize the scalar and gauge field as
\begin{eqnarray}
A_\mu=\Bigr[P_0(r),0,0,\frac{1}{e}(P(r)-n)\Bigr], \qquad \tilde\Phi(r)=\eta X(r)e^{in\varphi}\label{eq3-5}.
\end{eqnarray}
One  redefined $\bar\omega^2 \equiv-\frac{6\omega^2}{\kappa^2}$\cite{thooft6:2015} and $\Phi=\frac{1}{\omega}\tilde\Phi$.  We ignore, for the time being, fermion terms. The gauge covariant derivative is $D_\mu\Phi=\nabla_\mu\Phi+i\epsilon A_\mu\Phi$ and $F_{\mu\nu}$ the abelian field strength. We see that $\omega$ is rotated in the complex plane, necessary to make the integration over the dilaton technically identical to the integration over a conventional, renormalizable scalar field. The result is that one obtains an effective action for $\tilde g_{\mu\nu}$\cite{veltman:1974}. It is remarkable, as we shall see in section 4, that in our example of the spinning string spacetime, an exact complex solution is found for $\omega(r)$.

The cosmological constant $\Lambda$ could be ignored from the point of view of naturalness in order to avoid the inconceivable  fine-tuning. Putting $\Lambda$ zero increases the symmetry of the model.
This Lagrangian  is local conformal invariant under the transformation $\tilde g_{\mu\nu}\rightarrow\Omega^2 \tilde g_{\mu\nu}, \tilde \Phi \rightarrow \frac{1}{\Omega}\tilde \Phi$ and $\bar\omega\rightarrow \frac{1}{\Omega}\bar\omega$.

Varying the Lagrangian with respect to $\tilde g_{\mu\nu}, \tilde \Phi, \bar\omega$ and $A_\mu$, we obtain the equations

\begin{eqnarray}
\tilde G_{\mu\nu}=\frac{1}{(\bar\omega^2 +\tilde\Phi\tilde\Phi^*)}\Bigl(\tilde T_{\mu\nu}^{(\bar\omega)}+\tilde T_{\mu\nu}^{(\tilde\Phi,c)}+\tilde T_{\mu\nu}^{(A)}+\frac{1}{6}\tilde g_{\mu\nu}\Lambda\kappa^2\bar\omega^4
+\tilde g_{\mu\nu}V(\tilde\Phi,\bar\omega)\Bigr),\label{eq3-6}
\end{eqnarray}
\begin{eqnarray}
\tilde\nabla^\alpha \partial_\alpha\bar\omega -\frac{1}{6}\tilde R\bar\omega -\frac{\partial V}{\partial \bar\omega}-\frac{1}{9}\Lambda \kappa^2\bar\omega^3=0, \label{eq3-7}
\end{eqnarray}
\begin{eqnarray}
 \tilde D^\alpha \tilde D_\alpha\tilde\Phi-\frac{1}{6}\tilde R\tilde\Phi-\frac{\partial V}{\partial\tilde\Phi^*}=0,\qquad \tilde\nabla^\nu F_{\mu\nu}=\frac{i}{2}\epsilon \Bigl(\tilde\Phi ( \tilde D_\mu\tilde\Phi)^* -\tilde\Phi^*  \tilde D_\mu\tilde\Phi\Bigr),\label{eq3-8}
\end{eqnarray}
with
\begin{eqnarray}
\hspace{-0.5cm}
\tilde T_{\mu\nu}^{(A)}=F_{\mu\alpha}F_\nu^\alpha-\frac{1}{4}\tilde g_{\mu\nu}F_{\alpha\beta}F^{\alpha\beta},\label{eq3-9}
\end{eqnarray}
\begin{eqnarray}
\tilde T_{\mu\nu}^{(\tilde\Phi ,c)}=\Bigl(\tilde\nabla_\mu\partial_\nu\tilde\Phi\tilde\Phi^*-\tilde g_{\mu\nu}\tilde\nabla_\alpha\partial^\alpha\tilde\Phi\tilde\Phi^*\Bigr)\cr
-3\Bigl[ \tilde D_\mu\tilde\Phi( \tilde D_\nu\tilde\Phi)^*+( \tilde D_\mu\tilde\Phi)^* \tilde D_\nu\tilde\Phi
-\tilde g_{\mu\nu} \tilde D_\alpha\tilde\Phi( \tilde D^\alpha\tilde\Phi)^*\Bigl]\label{eq3-10}
\end{eqnarray}
and
\begin{eqnarray}
\hspace{-0.5cm}
\tilde T_{\mu\nu}^{(\bar\omega)}=\Bigl(\tilde\nabla_\mu\partial_\nu\bar\omega^2-\tilde g_{\mu\nu}\tilde\nabla_\alpha\partial^\alpha\bar\omega^2\Bigr)
-6\Bigl(\partial_\mu\bar \omega\partial_\nu\bar \omega-\frac{1}{2}\tilde g_{\mu\nu}\partial_\alpha\bar \omega\partial^\alpha\bar\omega)\Bigl).\label{eq3-11}
\end{eqnarray}
The covariant derivatives are taken with respect to $\tilde g_{\mu\nu}$.
Newton's constant reappears in the quadratic interaction term for the scalar field. One refers to the field $\bar\omega(r)$ as a dilaton field. A massive term in $V(\tilde\Phi,\bar\omega)$ will break the tracelessness of the energy momentum tensor,
a necessity for conformal invariance unless we would choose
\begin{eqnarray}
\frac{2}{3}V=\tilde \Phi^*\frac{\partial V}{\partial\tilde\Phi^*}+ \bar\omega\frac{\partial V}{\partial \bar\omega}.\label{eq3-12}
\end{eqnarray}
From the Bianchi identities we  obtain the additional equation
\begin{equation}
\frac{1}{6}V' =\tilde \Phi^{*'}\frac{\partial V}{\partial \tilde\Phi^*} +\bar\omega'\frac{\partial V}{\partial \bar\omega},\label{eq3-13}
\end{equation}
where a prime  represents $\frac{\partial}{\partial r}$.
\subsection{The interior of  the spinning cosmic string}\label{sec:level3c}

If we write out the field equations of section 3.1, we obtain (n=1 and $\Lambda=0$ for the moment)
\begin{equation}
\hspace{-2cm}
J''=J'\partial_r\Bigl[\log\Bigl(\frac{b}{\eta^2X^2+\bar\omega^2}\Bigr)\Bigr]-2\frac{P_0'(eJ P_0'+P')}{e(\eta^2X^2+\bar\omega^2)}
+12e^{2\mu}\frac{\eta^2eX^2P_0(eJP_0+P)}{\eta^2X^2+\bar\omega^2}\label{eq3-14},
\end{equation}
\begin{equation}
\hspace{-2cm}
\mu''=\frac{(J')^2}{2b^2}-\mu'\partial_r\Bigl[\log\Bigl(b(\eta^2X^2+\bar\omega^2)\Bigr)\Bigr]-\frac{(P'_0)^2}{\eta^2X^2+\bar\omega^2}
+6e^{2\mu}\frac{\eta^2 e^2 X^2P_0^2}{\eta^2X^2+\bar\omega^2}\label{eq3-15},
\end{equation}
\begin{eqnarray}
b''=\frac{(J')^2}{b}-b'\partial_r\Bigl[\log(\eta^2X^2+\bar\omega^2)\Bigr]-\frac{(eJP'_0+P')^2+e^2b^2(P_0')^2}{e^2 b(\eta^2X^2+\bar\omega^2)}\cr
+6\eta^2e^{2\mu}X^2\frac{P_0^2e^2(b^2+J^2)+P(2P_0Je+P)}{b(\eta^2X^2+\bar\omega^2)}\label{eq3-16},
\end{eqnarray}
\begin{equation}
X''=\frac{(J')^2X}{12b^2}-\frac{1}{3}X(\mu''+\frac{b''}{b})-\frac{b'X'}{b}+\frac{Xe^{2\mu}}{b^2}\Bigl((P_0Je+P)^2-e^2b^2P_0^2\Bigr)+e^{2\mu}\frac{dV}{dX}\label{eq3-17},
\end{equation}
\begin{equation}
P''=P'\Bigl(\frac{b'}{b}-J\frac{J'}{b^2}\Bigr)+eP'_0\Bigl(2J\frac{b'}{b}-J'\frac{(J^2+b^2)}{b^2}\Bigr)-\eta^2e^2e^{2\mu}X^2P\label{eq3-18},
\end{equation}
\begin{equation}
P_0''=P'_0\Bigl(J\frac{J'}{b^2}-\frac{b'}{b}\Bigr)+\frac{J'P'}{eb^2}-\eta^2e^2e^{2\mu}X^2P_0\label{eq3-19},
\end{equation}
\begin{figure}[h]
\centerline{
\includegraphics[width=4.8cm]{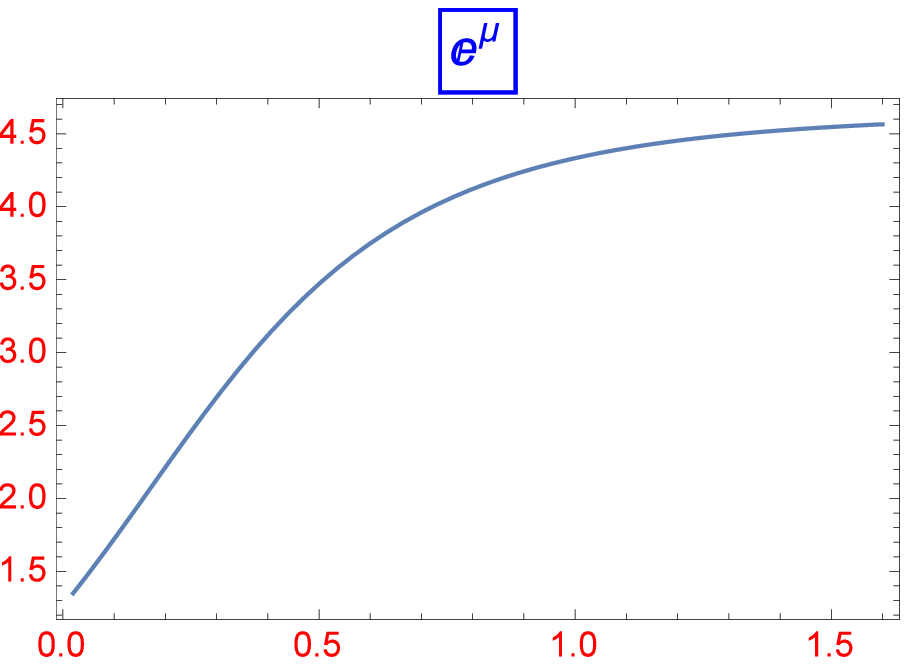}
\includegraphics[width=4.8cm]{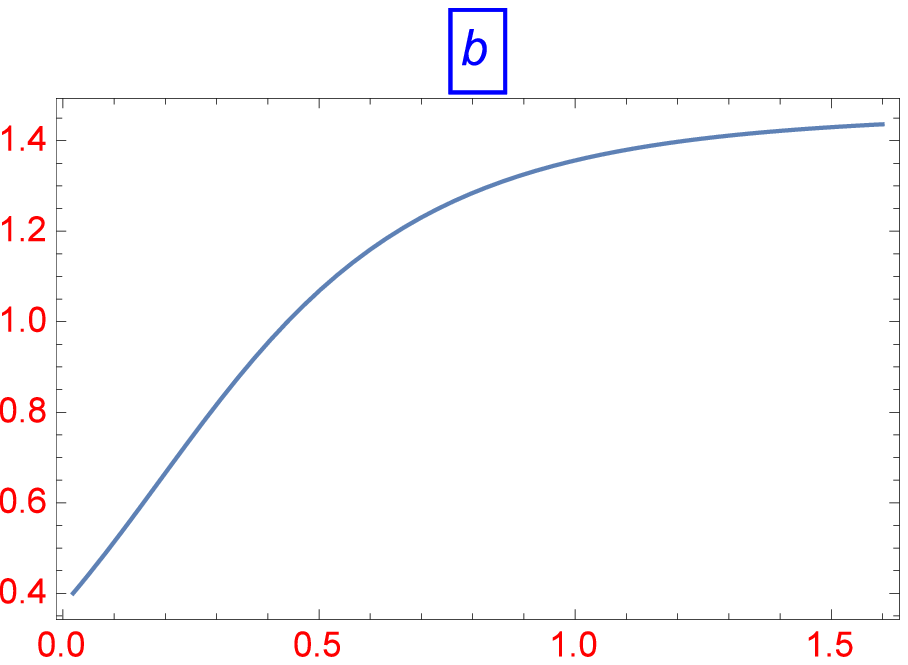}
\includegraphics[width=4.8cm]{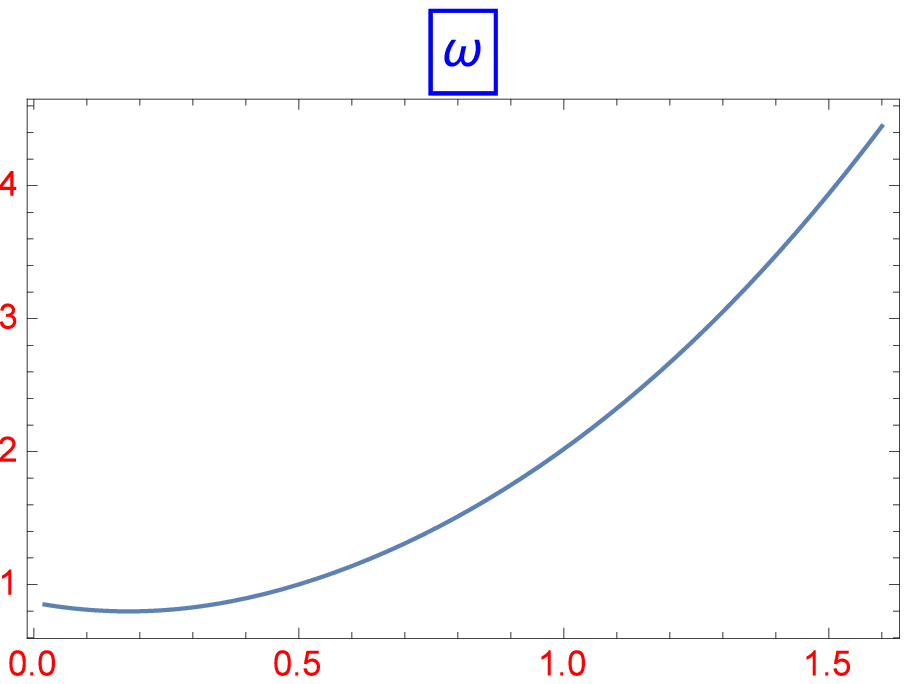}}
\centerline{
\includegraphics[width=4.8cm]{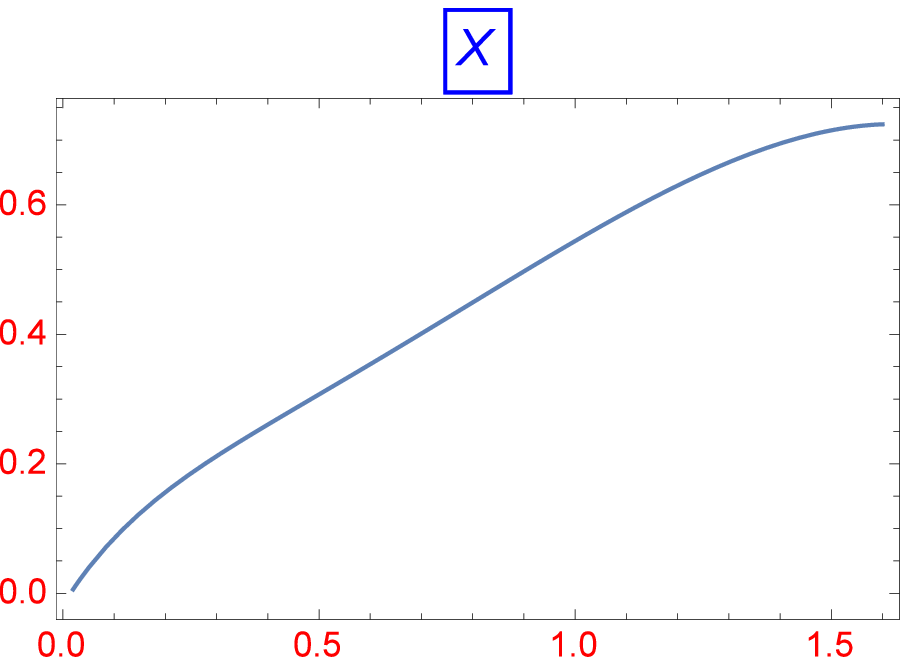}
\includegraphics[width=4.8cm]{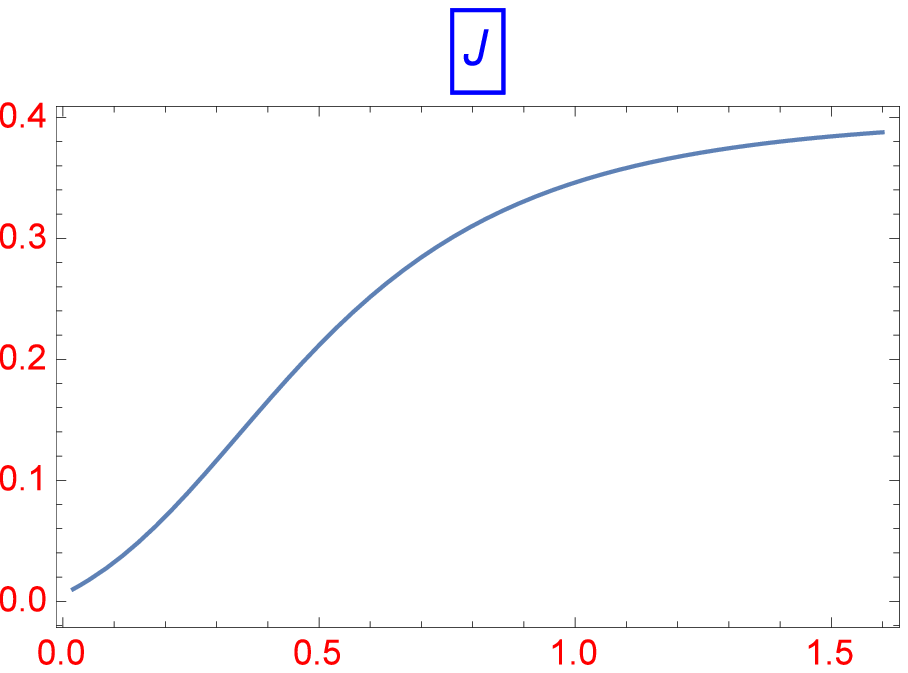}
\includegraphics[width=4.8cm]{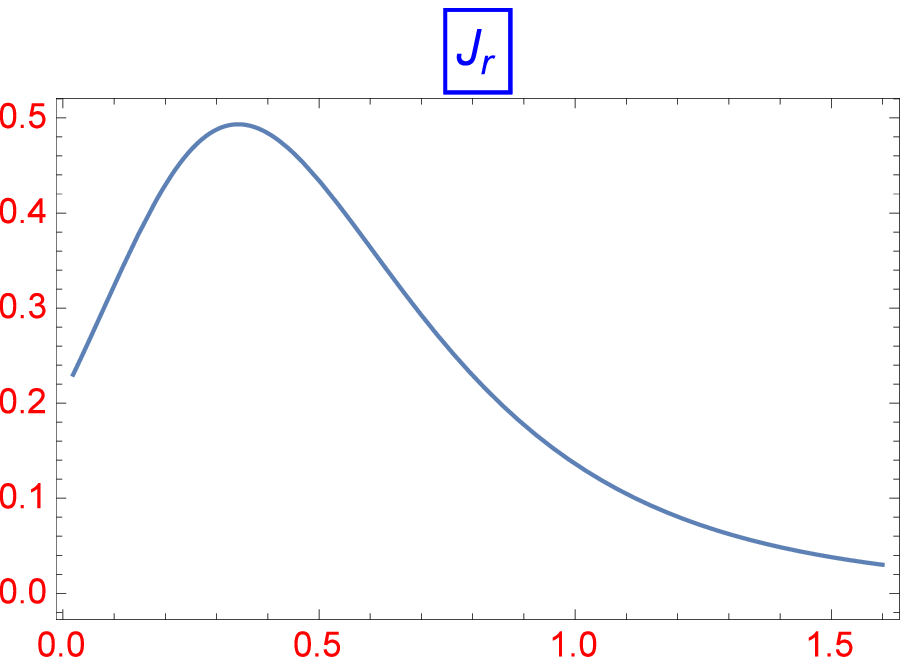}}
\centerline{
\includegraphics[width=4.8cm]{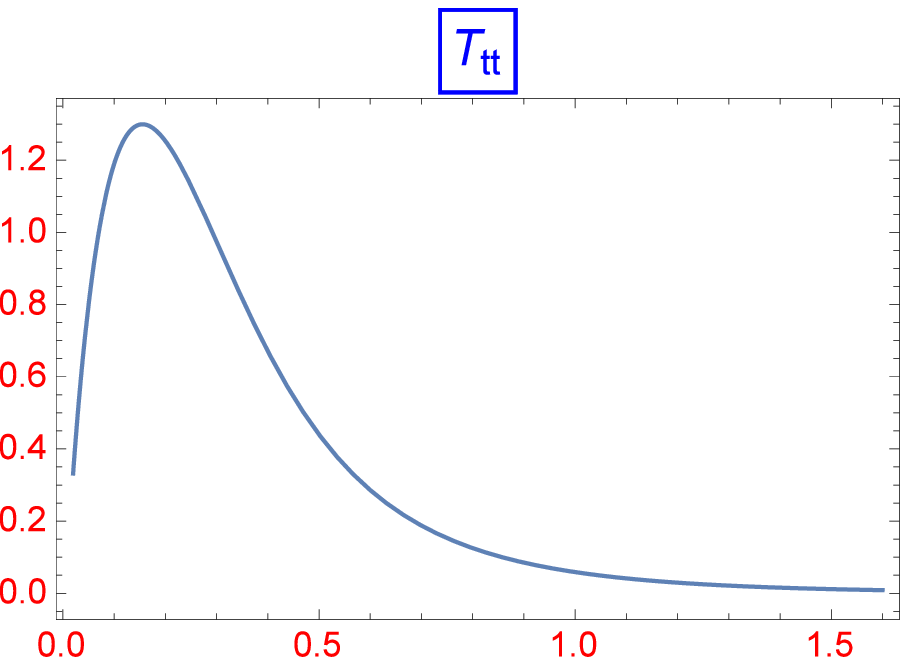}
\includegraphics[width=4.8cm]{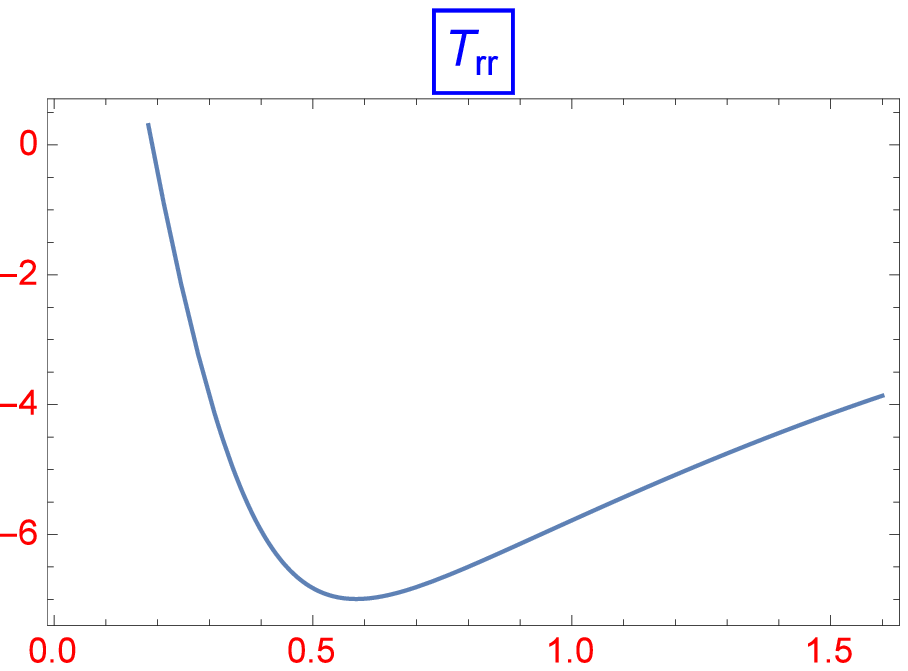}
\includegraphics[width=4.8cm]{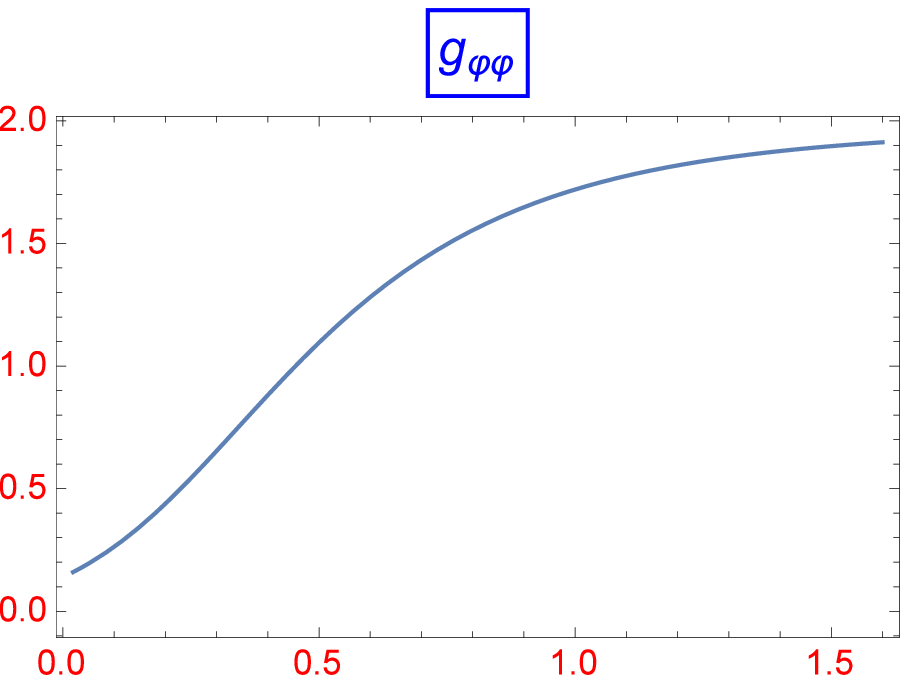}}
\caption{Plot of a typical solution in the case of global strings. The core of the string  is located at $r=r_C$ where $T_{tt}\approx 0$ and $X'\rightarrow0$. We observe that $J$ has the correct behavior.}
\end{figure}
\begin{equation}
\bar\omega''=\frac{\bar\omega(J')^2}{12b^2}-\frac{\bar\omega'b'}{b}
-\frac{1}{3}\bar\omega\Bigl(\mu''+\frac{b''}{b}\Bigr)+e^{2\mu}\frac{dV}{d\bar\omega}\label{eq3-20}.
\end{equation}
In the equation for $\bar\omega$ and $X$ one can of course eliminate the second order derivative terms on the right-hand side.
Eq.(\ref{eq3-20}) is the dilaton equation. In the numerical solution we will use the $\bar\omega$ equation from the Einstein equations and will use  Eq.(\ref{eq3-20}) as constraint.
Further, there is again a constraint equation for $J'$.
For global strings, i.e., $P=P_0=0$, we have from Eq.(\ref{eq3-14}) a spin-mass relation
\begin{equation}
J=cons\int\frac{b}{\eta^2X^2+\bar\omega^2}dr\label{eq3-21}.
\end{equation}
The components $T_{tt}$ and $T_{rr}$  of the energy-momentum tensor  become
\begin{eqnarray}
T_{tt}=-\frac{3}{4b^2}(J')^2+\frac{\mu' b'}{b}+(\mu'+\partial_r \log b)\partial_r\Bigl(\log(\eta^2X^2+\bar\omega^2)\Bigr)\label{3-22}
\end{eqnarray}
\begin{eqnarray}
T_{zz}=\frac{3}{4b^2}(J')^2-\frac{\mu' b'}{b}-\partial_r \log b\partial_r\Bigl(\log(\eta^2X^2+\bar\omega^2)\Bigr)\label{eq3-23}.
\end{eqnarray}
Note that $T_{tt}=-T_{zz}$ for $\mu=0$.
\subsection{A numerical solution}\label{sec:level3c}
A numerical solution in case of global stings is plotted in figure 1.
At the core of the string, it is always possible to match this solution on the exterior, specially the component $J$, which causes in standard GRT problems.
From the behavior of the energy momentum tensor components $T_{tt}$ and $T_{rr}$ , we observe that in this example the strong energy condition is fulfilled by suitable initial conditions.
Further, it is always possible to make $g_{\varphi\varphi}=b^2-J^2$ positive for $r$ not to close to $r=0$ and for suitable initial conditions.
\section{The exterior vacuum solution}\label{sec3}
\subsection{An exact solution without CTC's}\label{sec3}
For the exterior vacuum, the  field equations become (with cosmological constant)
\begin{equation}
J''=J'\Bigl(\frac{b'}{b}-2\frac{\bar\omega'}{\bar\omega}\Bigr),\label{eq4-1}
\end{equation}
\begin{equation}
b''=\frac{1}{b}(J')^2-\frac{2}{\bar\omega}b'\bar\omega'\label{eq4-2},
\end{equation}
\begin{equation}
\mu''=\frac{1}{2b^2}(J')^2-\mu'\Bigl(\frac{b'}{b}+2\frac{\bar\omega'}{\bar\omega}\Bigr)\label{eq4-3},
\end{equation}
\begin{equation}
\bar\omega''=-\frac{3\bar\omega}{8b^2}(J')^2+\frac{(\bar\omega')^2}{2\bar\omega}+\frac{1}{2}\mu'\Bigl(\frac{\bar\omega b'}{b}+2\bar\omega'\Bigr)+\frac{1}{2}\Lambda\kappa^2e^{2\mu}\omega^3\label{eq4-4}.
\end{equation}
Further, we have the constraint equations
\begin{equation}
( J')^2=-\frac{12b^2}{\bar\omega^2}(\bar\omega')^2-4b\mu'(b'+\frac{2b}{\bar\omega}\bar\omega')-\frac{8b}{\bar\omega}b'\bar\omega'+4\Lambda\kappa^2\frac{b^2}{\omega^2}e^{2\mu}\label{eq4-5}.
\end{equation}
These equations can be solved exactly for $\Lambda =0$. We  immediately observe that Eq.(\ref{eq4-1}) leads to
\begin{equation}
J(r)=cons.\int\frac{b(r)}{\bar\omega(r)^2} dr\label{eq4-6},
\end{equation}
a kind of "spin-mass" relation. In the interior case, the integral will also contain the scalar and gauge field ( see also Eq. (\ref{eq3-21})).
The most interesting Ricci-flat solution is
\begin{eqnarray}
\hspace{-0.4cm}
\mu(r)=c_1r+c_2-\log(\sqrt{c_4r+c_5}),\qquad b(r)=\frac{c_3}{2c_4r+2c_5},\qquad \bar\omega(r)=\sqrt{2c_4r+2c_5},\cr J(r)=c_6\pm\frac{c_3}{2c_4r+2c_5}\label{eq4-7}.
\end{eqnarray}
If we calculate the metric component $\tilde g_{\varphi\varphi}=b(r)^2-J(r)^2$  or $g_{\varphi\varphi}=\bar\omega(r)^2\tilde g_{\mu\nu}$, we will encounter a CTC for $r=\frac{\pm c_3-c_5c_6}{c_4c_6}$, which can be pushed to $\pm\infty$ by suitable choices of the parameters. The behavior of $J(r)$ has asymptotically the correct form.
The spacetime $g_{\mu\nu}$ is Ricci flat, while $\tilde g_{\mu\nu}$ is not.
We could also consider the inverse transformation $\tilde g_{\mu\nu}=\frac{1}{\omega^2} g_{\mu\nu}$ in order to generate from flat spacetime a non-flat spacetime by the dilaton field. In figure 2 we plotted a numerical solution of the Eq.(\ref{eq4-1})-(\ref{eq4-4}). It confirms the correct behavior of $J$ and the dilaton field as scale factor.
\begin{figure}[h]
\centerline{
\includegraphics[width=4.8cm]{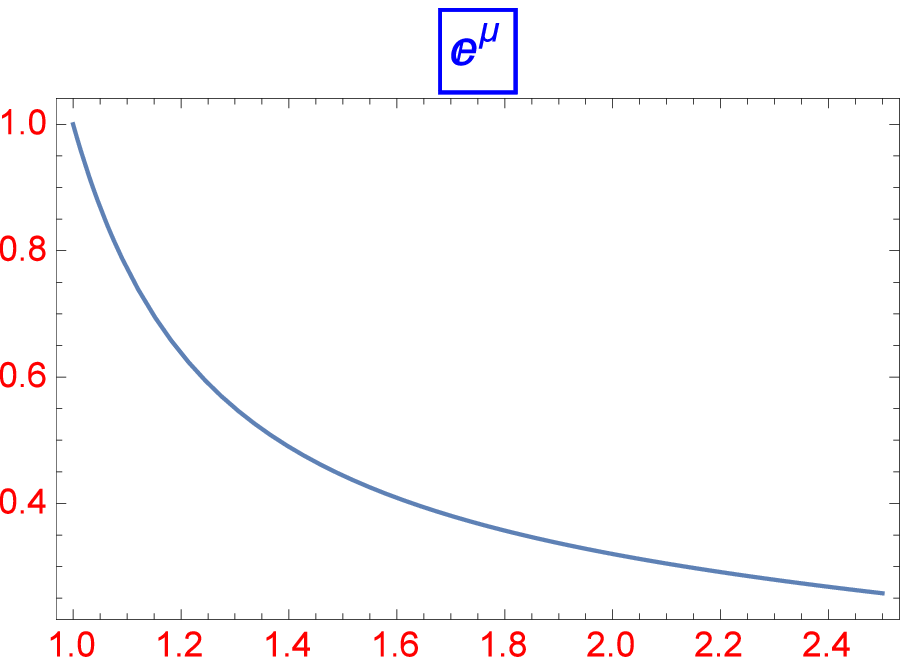}
\includegraphics[width=4.8cm]{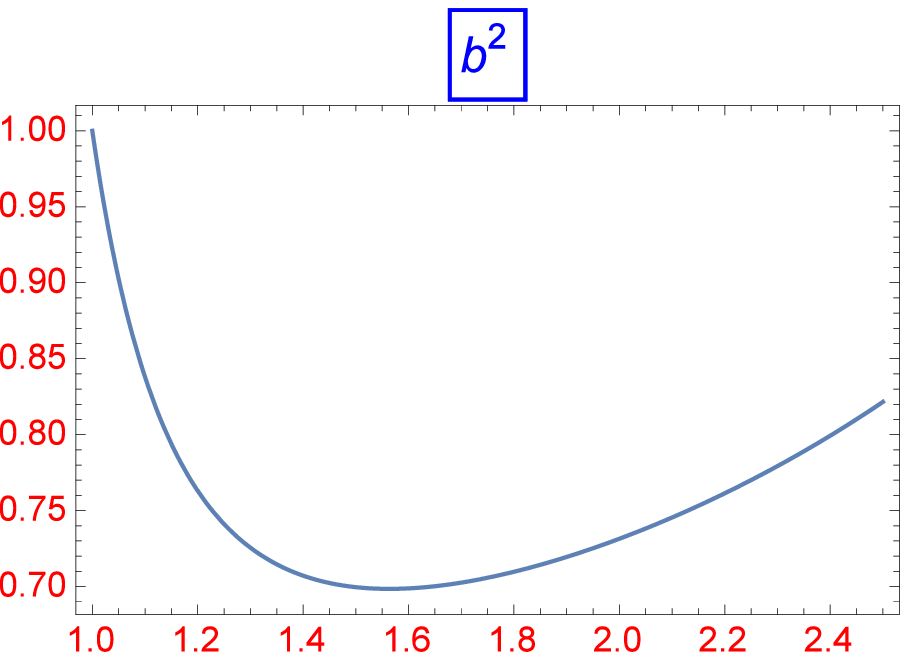}
\includegraphics[width=4.8cm]{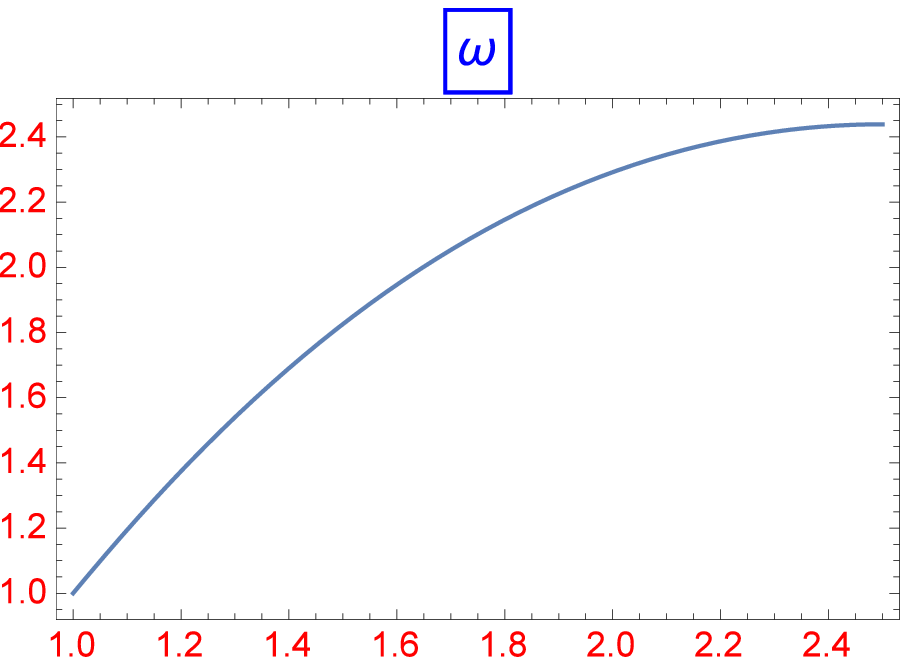}}
\centerline{
\includegraphics[width=4.8cm]{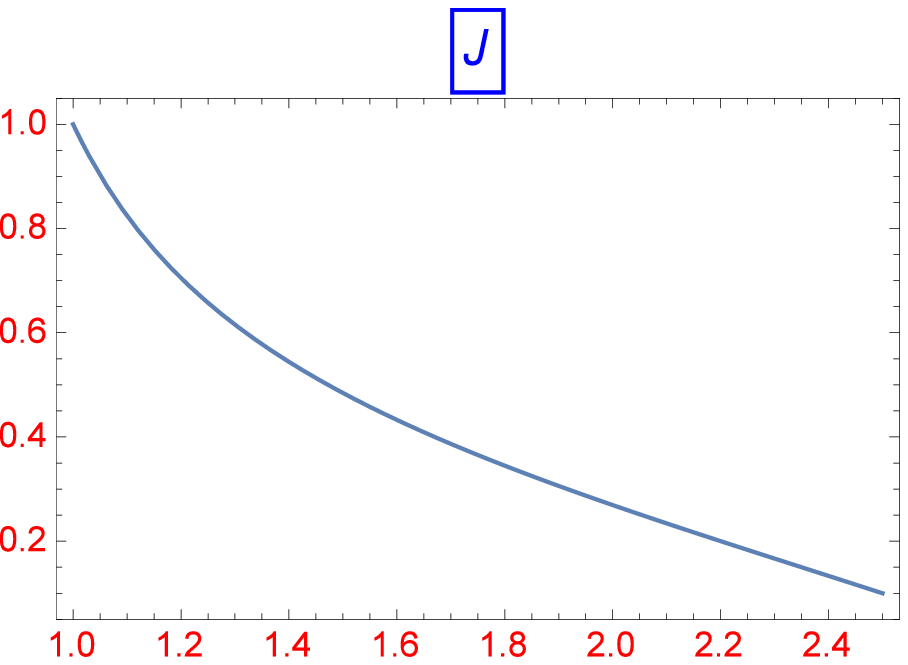}
\includegraphics[width=4.8cm]{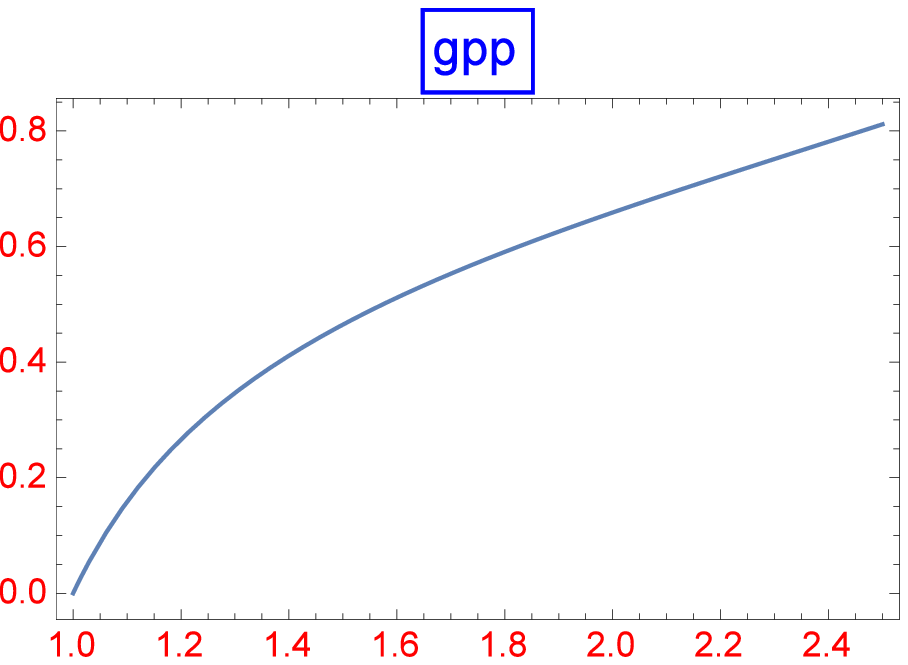}
\includegraphics[width=4.8cm]{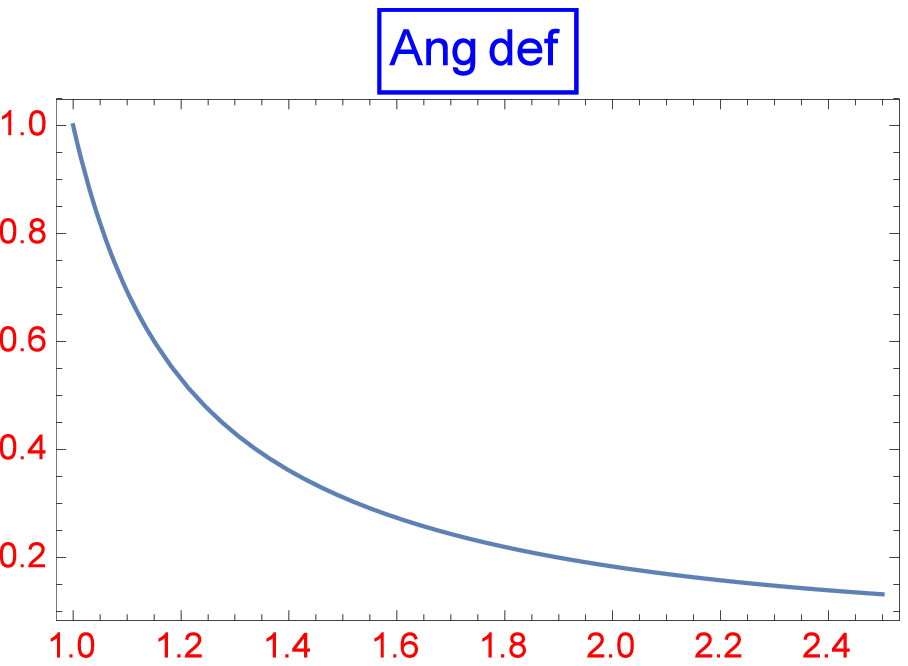}}
\caption{Plot of a typical exterior vacuum solution with the correct asymptotic behavior of the spin parameter $J$. The core of the string is located close to $r=1$.}
\end{figure}
Outside the cosmic string one should experience an angle deficit $\Delta \phi$, which is not the zero-thickness  limit $\Delta \phi \approx \mu$ in the  weak field approximation\cite{futamase:1988}, with $\mu$ the mass per unit length. Of course, the relation between the angle deficit and mass depends crucially on the energy-momentum tensor, i.e., the initial condition on the core of the string.
In figure 2 we also plotted $\frac{b^2}{r^2}$, an indication of the angle deficit.
In figure 3 we plotted another solution, with a cosmological constant.
\begin{figure}[h]
\centerline{
\includegraphics[width=4.8cm]{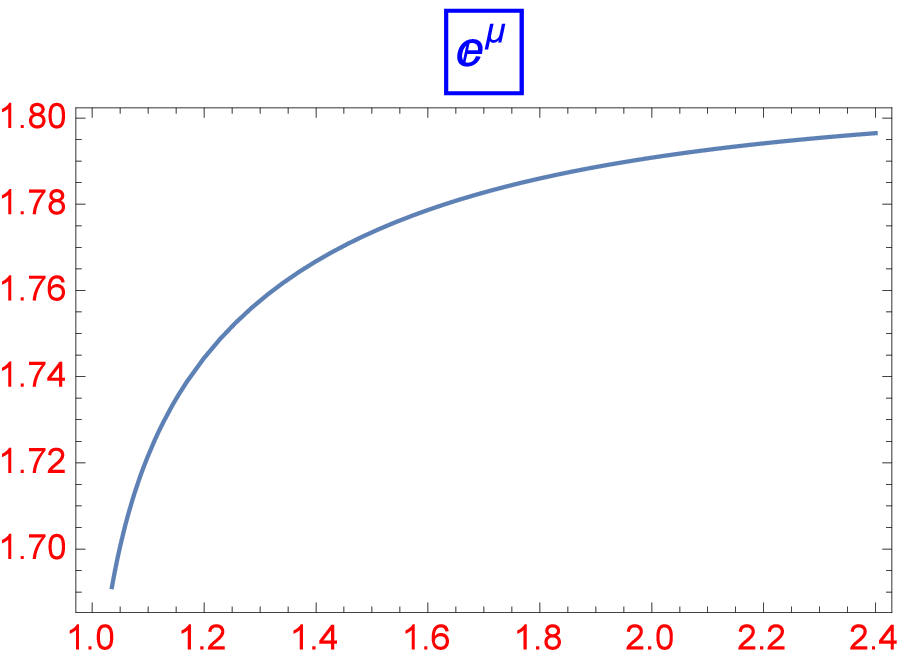}
\includegraphics[width=4.8cm]{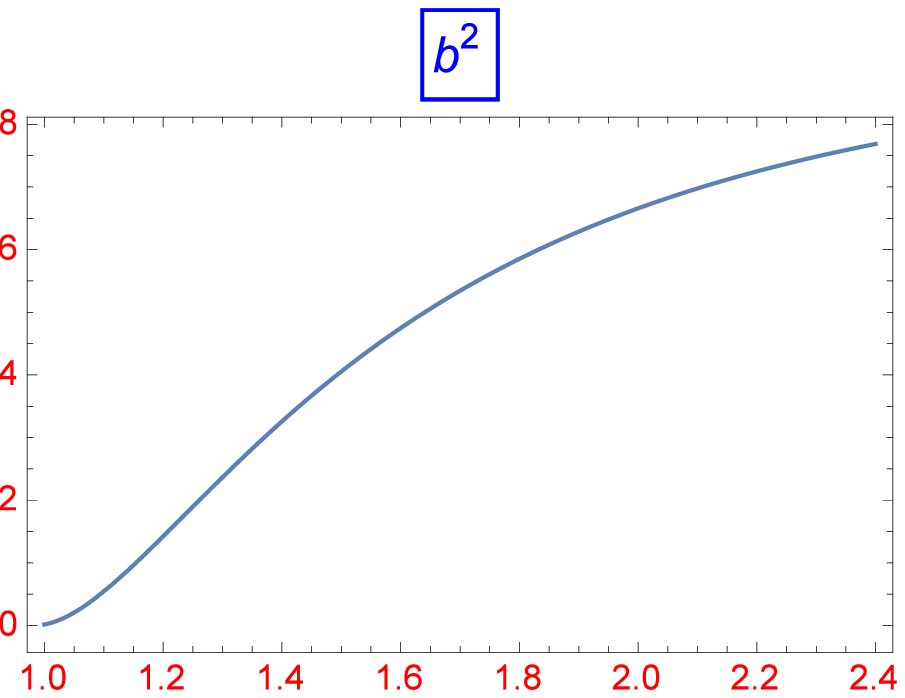}
\includegraphics[width=4.8cm]{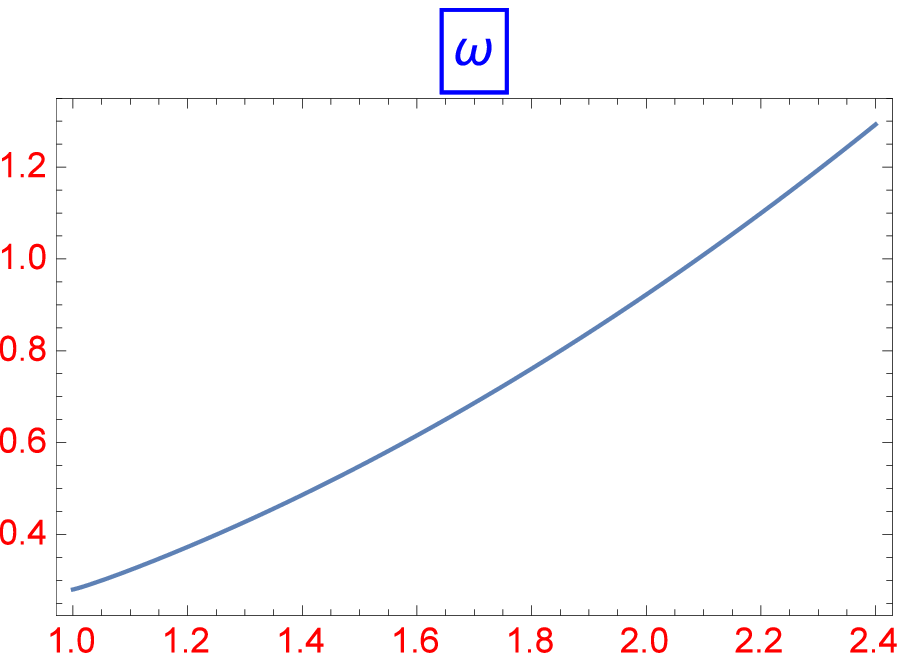}}
\centerline{
\includegraphics[width=4.8cm]{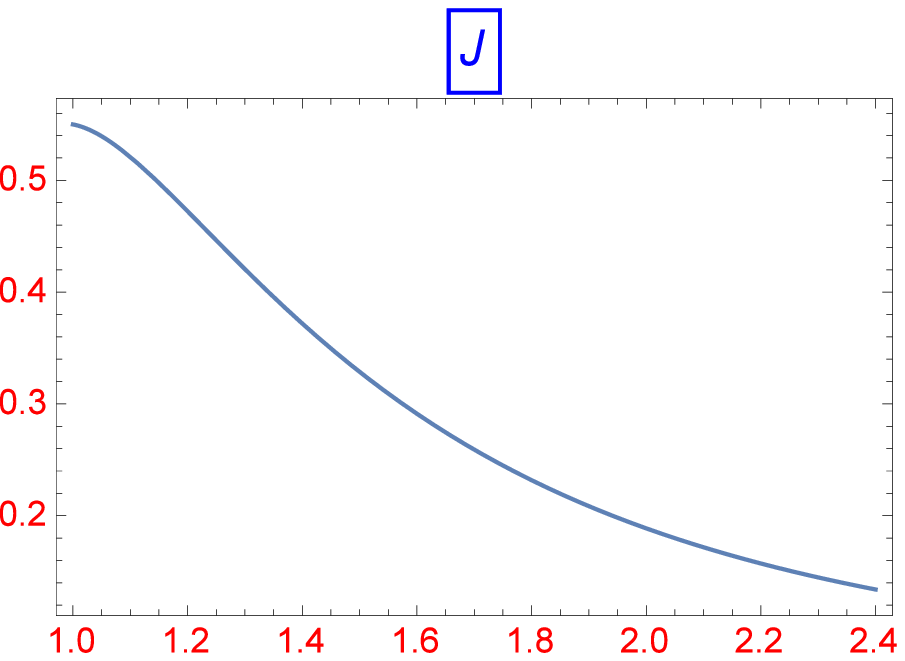}
\includegraphics[width=4.8cm]{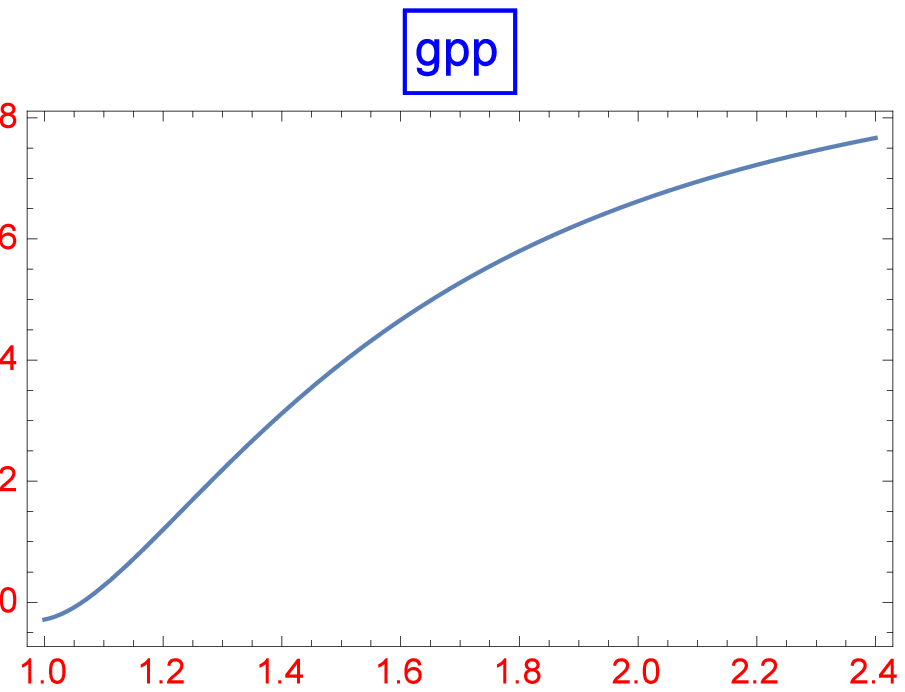}
\includegraphics[width=4.8cm]{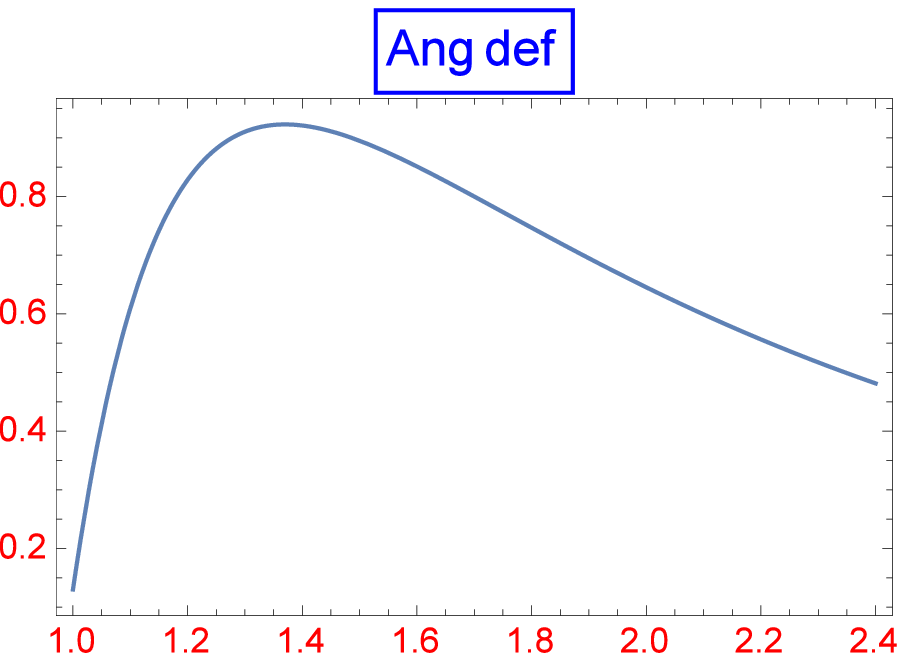}}
\caption{Plot of a typical exterior vacuum solution with a cosmological constant. $J$ shows the correct behavior close to the core as well asymptotically.  }
\end{figure}

Another asymptotic solution to this set of equations that possesses the correct behavior for $J(r)$ is
\begin{eqnarray}
\mu(r)=-\frac{5}{4}\log(\bar{\omega}(r))+c_1,\qquad b(r)=\frac{c_2}{\gamma^{2/3}\sqrt{c_3r+c_4}}, \qquad \bar{\omega}(r)=\gamma(c_3r+c_4)^{3/4}, \cr
 J(r)=c_6- \frac{c_2}{\gamma^{8/3}c_3(c_3r+c_4)},\label{eq4-8}
\end{eqnarray}
with $\gamma=\frac{108^{3/4}}{27}$ or $\gamma=\frac{108^{3/4}}{27}I$ . By setting the integration constant $c_6=-\frac{1}{4}\gamma^{4/3} c_2c_3$, the CTC occurs at $r=\frac{4-\gamma^4c_3^2c_4}{\gamma^4c_3^3}$. The curvature of this solution (for $g_{\mu\nu}$) is
\begin{equation}
R=\frac{e^{-2c_1}}{2\gamma^{7/2}(c_3r+c_4)^{21/8}},\label{eq4-9}
\end{equation}
which is independent of the integration variable $c_6$. There is a curvature singularity here at $r_0=-c_4/c_3$ for particular values of the parameters. Inspection of the metric components, for example $g_{tt}=-\bar{\omega}^2$, reveals that the spacetime is nonsingular if the metric is real.
\subsection{A local observer: no WEC violation}\label{sec5}
Let us consider a local  orthonormal tetrad frame $\hat\Theta^\mu: \hat\Theta^t =dt-Jd\varphi, \hat\Theta^r=e^\mu dr, \hat\Theta^z= e^\mu dz, \hat\Theta^\varphi =b d\varphi$.
For a timelike four-velocity vector field $U_{\hat\mu}=\frac{1}{\epsilon}[1,0,\alpha,\beta]$ , with $\alpha^2+\beta^2 =1-\epsilon^2$, we have  the local energy density measured
by the observer moving at constant  $r=r_s$  (for $\mu =0$ )
\begin{eqnarray}
\epsilon^2 G^{\hat\mu\hat\nu}U_{\hat\mu} U_{\hat\nu}=-\frac{\beta}{b}J''-\frac{\beta^2+\epsilon^2}{b}b''+\frac{2\beta^2+\epsilon^2+2}{4b^2}(J')^2\label{eq4-10}.
\end{eqnarray}
This  expression is positive  for all $\alpha , \beta$ and $\epsilon$ for  the physically acceptable behavior of $J''<0$ and $b''<0$. If we substitute the field equations into Eq.(\ref{eq4-10}), we obtain
\begin{equation}
\frac{\partial_r\Bigl(\log(\eta^2 X^2+\omega^2)\Bigr)}{b}\Bigl((\beta^2+\alpha^2)b'+\beta J'\Bigr)+\frac{2\alpha^2-\epsilon^2}{4b^2}(J')^2\label{eq4-11},
\end{equation}
which is also positive for suitable physically acceptable behavior of the derivatives of $J, b, X, \omega$ and $\epsilon^2 <2\alpha^2$.
\subsection{\label{sec:level3e} The "unphysical" metric $\tilde g_{\mu\nu}$ related to the $(2+1)$ Kerr spacetime }
We can compare the conformal solution with the standard solution of GRT. The solution of $\tilde g_{\mu\nu}$ in the non-conformal case is easily found:  $J_0=$ constant, $\mu =0$ and $b(r)=pr+q$ ($p$ and $q$ constants), i.e., a Kerr-type solution with two conserved parameters $(p,J)$ representing the the mass and angular momentum of a spinning point source\cite{deser2:1992}. This solution is locally flat, but possesses manifestly CTC's and requires unphysical sources.
The CTC arises if $b(r)=pr+q <J_0$. For a standard cosmic string we can write $ b=(1-4G\mu)r$, with $\mu$ the mass density of the string. So the criterium is $ r_s <\frac{J_0}{1-4G\mu}$, with $\mu<\frac{1}{4G}$.
The physical question is, can the source of the string be confined within a small enough region in order to satisfy this criterium.
In general, this criterium will not be fulfilled. It was found\cite{slagter5:2015} in a numerical approximation, that the formation of CTC's outside the core of a global string is  unlikely.
One can change the coordinates in order to hide the presence of the string and make the spacetime appear locally Minkowski, i.e., $t^* =t-J_0 \varphi$ and $\varphi^* =(1-4G\mu)\varphi$. The new time variable, however, jumps by $J_0$ whenever the string is circumnavigated.  Needless to say, a CTC is then every where. See for example the discussion of Deser, et al.\cite{deser3:1992}.
Further, the spacetime is conical, i.e., $0<\varphi^*<2\pi(1-4G\mu)$.
We also observe that the $z$-coordinate is a dummy coordinate. There is no structure in the $z$-direction, hence one can suppress that direction. The resulting planar (2+1)-dimensional gravity spacetime
\begin{equation}
ds^2=-(dt-Jd\varphi)^2+b^2d\varphi^2+dr^2\label{eq4-12}
\end{equation}
was  intensively studied\cite{thooft2:1992,thooft3:1993}, in relation with quantizing the model.
The Einstein tensor density has a singularity describing a massive spinning point-source or "cosmon", $T^{tt}\sim4G\mu\delta^2(r), T^{ti}\sim J\epsilon^{ij}\partial_j\delta^2(r)$. When we "lift up" the $(2+1)$-dimensional spacetime to $(3+1)$-dimensional spacetime, then this viewpoint is no longer tenable: an infinite thin cosmic string ( an infinite line mass) is physical not acceptable. It must have a boundary.
The one-parameter Minkowski metric minus a wedge is sometimes describes as a thin-string approximation and considered in this context as the gravitational field of an infinite thin wire with distributional stress energy if the radial stress is negligible compared with the energy density\cite{israel:1977}. This is impossible for cosmic strings. Geroch and Traschen\cite{geroch:1987} show that the  metric of an infinite thin string cannot be regular can cannot assign a distributional stress-energy tensor to Minkowski minus a wedge.
Problems will also arise at the boundary with the interior: the WEC is violated\cite{janca:2007}.
The exterior solution of Eq(\ref{eq4-1}), found by van Stockum\cite{stockum:1937} and analyzed by Krisch\cite{Krisch:2003} cannot represent a realistic solution, because $J(r)\sim r^2$.
\section{\label{sec:level3}The warped 5D connection and holography}
Brane world models with non-compact extra dimensions are interesting modified gravity models. In this scenario, our 4D world is described by a brane $({\cal M},{^{(4)\!}g}_{\mu\nu})$ in a 5D spacetime $({\cal V},{^{(5)\!}g}_{\mu\nu})$.
All Standard Model fields are confined to the brane while gravity is free to access the bulk.
At low energy, gravity is localized on the brane and  GRT is recovered. The long-standing hierarchy problem would be solved, i.e., why there is such a large gap between the electroweak scale $M_{ew}\sim 10^{3}GeV$  and the  Planck scale $M_{pl}\sim 10^{19}GeV$. In the warped model of Randall-Sundrum (RS)\cite{ran1:1999,ran2:1999} with one large extra dimension L, the observed 3 dimensions are protected from the large extra dimension at low energies, by curvature rather than straightforward compactification. One speaks of "warped compactification".
An observer confined to the brane will experience gravity modified by the very embedding itself. The massless 5D graviton decomposes into a massless 4D graviton and an infinite tower of massive 4D spin 2 modes.
In the brane world picture it is possible that the 4D Planck scale is not fundamental, but only an effective scale. For a dimension of $L\sim 1$ mm, the fundamental Planck scale can be of the order of $M_{ew}$.
In a former study\cite{slagter2:2016} we considered the 5D warped axially symmetric spacetime, where the bulk spacetime is empty with only bulk cosmological constant $\Lambda_5$
\begin{equation}
ds^2_5={\cal W}({\bf x},y)^2{^{(4)\!}g}_{\mu\nu}({\bf x}) +\Gamma(y)^2dy^2\label{eq5-1}
\end{equation}
The extra non-compact dimension is $y$. Due to the "warp-factor" ${\cal W}$, it becomes very small as seen from the brane.
${\cal W}$ can be solve exactly from the 5D Einstein equations
\begin{equation}
{^{(5)\!}G}_{\mu\nu}=-\Lambda_5{^{(5)\!}g}_{\mu\nu}\label{eq5-2}
\end{equation}
resulting in
\begin{equation}
{\cal W}=\pm \frac{1}{\sqrt{\tau r}}e^{\sqrt{-\frac{1}{6}\Lambda_5}(y-y_0)}\sqrt{(d_1e^{\sqrt{2\tau}t}-d_2e^{-\sqrt{2\tau}t}  )(d_3e^{\sqrt{2\tau}r}-d_4e^{-\sqrt{2\tau}r})}\label{eq5-3}
\end{equation}
We recognize the RS warp factor, i.e., the $y$-dependent term. The second term is a scale factor from the bulk equations which has its reflection on the brane.
It turns out\cite{shir:2000} that the effective metric ${^{(4)\!}g}_{\mu\nu}$ must be solved from the effective 4D Einstein equations
\begin{equation}
{^{(4)\!}G}_{\mu\nu}=-\Lambda_{eff}{^{(4)\!}g}_{\mu\nu}+\kappa_4^2 {^{(4)}T}_{\mu\nu}+\kappa_5^4{\cal S}_{\mu\nu}-{\cal E}_{\mu\nu}\label{eq5-4}
\end{equation}
where   ${\cal S}_{\mu\nu}$ is the quadratic term in the energy momentum tensor arising from the extrinsic curvature terms and ${\cal E}_{\mu\nu}$ the part of the 5D Weyl tensor that carries information of the gravitational field outside the brane. $\Lambda_{eff}$ represents the effective cosmological constant on the brane. To get a sense of the relevance of these two terms in Eq.(\ref{eq5-4}) in cosmological context, we refer to Mannheim\cite{mannheim:2017}.
His model could  open a way to explain dark matter and dark energy.
From the junction conditions, one obtains $\Lambda_{eff}=\frac{1}{2}(\Lambda_5+\kappa_4^2\Lambda_4)$ and $\kappa_4^2=\frac{1}{6}\kappa_5^4\Lambda_4$. For the RS fine-tuning we have $\Lambda_4^2 \kappa_5^4=-6\Lambda_5$, which means that $\Lambda_{eff}=0$.
Newton's constant is $G_N\equiv8\pi\kappa_4^2$. Einstein's gravity is recovered by $\kappa_5\rightarrow 0$, while keeping $G_n$ finite.
The contribution of the warp factor ${\cal W}$ to the 4D manifold can be seen as a "holographic" effect. This is comparable with the AdS/CFT correspondence, in which the classical dynamics of the 5D model are equivalent to the quantum dynamics of a conformal field theory on the boundary.
It must be noted, that the ${\bf x}$-dependent part of the warp factor in this model can be seen, from the cosmological point of view, as a time-dependent "scale" factor. It determines the dynamics of the brane. However, on small scales (as we discussed in section 3 for the 4D spacetime), it can be interpreted as a dilaton field coming from the bulk. Its role is not quite clear, when one  enters the quantum scale of the model.
If we calculate the trace of the Einstein equations Eq.(\ref{eq5-4}), one obtains
\begin{equation}
\frac{1}{\bar\omega^2 +X^2}\Bigl[16\kappa_4^2\beta\eta^2X^2\bar\omega^2-\kappa_5^4n^4\Bigl(\frac{(\partial_r P)^2-(\partial_t P)^2}{r^2e^2}\Bigr)^2 e^{8\bar\psi -4\bar\gamma}\Bigr]\label{eq5-5}
\end{equation}
with $n$ the winding number (multiplicity) of the scalar field and where we used for the potential $V(\tilde\Phi\bar\omega)=\frac{1}{8}\beta\eta^2\kappa_4^2\tilde\Phi\tilde\Phi\bar\omega^2$. We observe, as expected, that the first term breaks the tracelessness. The second term comes from the quadratic term in the energy momentum tensor, ${\cal S}_{\mu\nu}$ and becomes important for higher energy scales. It could solve the trace anomaly at some scale.
If we write out the 5D Einstein equations Eq.(\ref{eq5-2}) for the metric
\begin{equation}
ds^2=\omega^2\Bigl[-(dt-Jd\varphi)^2+b^2d\varphi^2+e^{2\mu}(dr^2+dz^2)\Bigr]+dy^2\label{eq5-6}
\end{equation}
with $\Lambda_5$ the bulk cosmological constant and y the extra dimension, we obtain exactly the system of equations of section (4.1).
It is remarkable that the exterior solution in the case of the spinning string in conformal gravity, is identical with the solution which will follow from the warped 5D Einstein equations.
It is a nice example of the equivalency of conformal invariance with the warped spacetime of one dimension higher.
\section{Conclusions}\label{sec5}
In  standard GRT , the issue of spinning string solutions has a long history. Famous are the van Stockum and Gott-Hiscock solutions\cite{Islam:1985,stockum:1937,hiscock:1985}. However, there are serious problems in these models. The weak energy condition is violated and it is hard to match the interior on the exterior solution\cite{janca:2007,Krisch:2003}.
One can easily verify that on the spacetime of Eq.(\ref{eq2-1}) in the standard general relativistic situation,  $f=0$ and only global string solutions are possible, i.e., $P=P_0=0$. In the exterior one finds that $J$ is a constant and it is troublesome to match smoothly $J$ at the boundary.

In our conformal invariant model these problems and restrictions  don't exist. It describes an example how the Ricci-flat spacetimes could generate curvature by a suitable dilaton field $\omega$ and gauge freedom $\Omega$.
The conformal component of the metric field is treated as a dilaton field $\omega$ on an equal footing as the scalar field. By demanding regularity of the action, no problems will emerge when $\omega\rightarrow0$.
For the exterior vacuum,  exact (Ricci-flat) solutions are found with the correct asymptotic features which can be matched on the numerical interior solution.
For global cosmic strings, the existence of CTC's can be avoided or pushed to infinity by suitable values of the integration constants. These constants can be used to fix the parameters of the cosmic string by the smooth matching of the solutions at the boundary. There seems to be no problems in order to fulfil the weak energy condition.
The numerical interior solution for gauge strings is harder to find due to the coupling of the differential equation for $J$ with the other field equations.

Our world, however, is not scale invariant, so the exact local conformal symmetry must be broken spontaneously. This means that we need additional field transformations on the vacuum spacetime, since $R$ transforms as
$R\rightarrow \frac{1}{\Omega^2}R-\frac{6}{\Omega^3}\nabla^\mu\partial_\mu\Omega$.
In order to obtain an effective conformal invariant and finite  theory, many problems must be overcome, such as  unitarity violation and conformal anomalies\cite{thooft0:2010}.
In canonical gravity, quantum amplitudes are obtained by integrating the action over all components of the metric $g_{\mu\nu}$. In conformal dilaton gravity, the integration is first done over $\omega$ and
then over $\tilde g_{\mu\nu}$ and imposing constraints on $\tilde g_{\mu\nu}$ and matter fields (including fermionic fields).
The effective action will then describe a conformally invariant theory both for gravity and matter. It could even be finite and renormalizable. By avoiding anomalies one can generate constraints which will determine the physical constants such as the cosmological constant. So the spontaneously broken CI could fix all the parameters.
In our investigated conformal invariant model of a spinning source, we find that $\omega$ and $\tilde g_{\mu\nu}$ can easily be found.
Our result  could be a new possible indication that local conformal invariance and spontaneously broken in the vacuum, can be a promising method for studying quantum effects in GRT, as was found in many other studies\cite{thooft4:2011,mannheim:2017,oda:2015}.

\end{document}